# Defective hierarchical porous copper-based metal-organic frameworks synthesised via facile acid etching strategy

Huan V. Doan,[1,2] Asel Sartbaeva,[3] Jean-Charles Eloi,[4] Sean Davis[5] and Valeska P. Ting[*1]

[1]Department of Mechanical Engineering, University of Bristol, Bristol BS8 1TH, UK.

[2]Department of Oil Refining and Petrochemistry, Faculty of Oil and Gas, Hanoi University of Mining and Geology, Duc Thang, Bac Tu Liem, Hanoi, Vietnam.

[3]Department of Chemistry, University of Bath, Claverton Down, Bath, BA2 7AY, UK.

[4]Chemical Imaging Facility, School of Chemistry, University of Bristol, Bristol BS8 1TS, UK.

[5]School of Chemistry, University of Bristol, Bristol BS8 1TS, UK.

*e-mail: v.ting@bristol.ac.uk

## Abstract

Introducing hierarchical pore structure to microporous materials such as metal-organic frameworks (MOFs) can be beneficial for reactions where the rate of reaction is limited by low rates of diffusion or high pressure drop. This advantageous pore structure can be obtained by defect formation, mostly via post-synthetic acid etching, which has been studied extensively on water-stable MOFs. Here we show that a water-unstable HKUST-1 MOF can also be modified in a corresponding manner by using phosphoric acid as a size-selective etching agent and a mixture of dimethyl sulfoxide and methanol as a dilute solvent. Interestingly, we demonstrate that the etching process which is time- and acidity- dependent, can result in formation of defective HKUST-1 with extra interconnected hexagonal macropores without compromising on the bulk crystallinity. These findings suggest an intelligent scalable synthetic method for formation of hierarchical porosity in MOFs that are prone to hydrolysis, for improved molecular accessibility and diffusion for catalysis.

## Introduction

Metal-organic frameworks (MOFs) are a new class of crystalline porous solids which can be self-assembled from an abundance of inorganic ions/clusters and a variety of organic linkers/ligands, thus possessing tailorable chemical and structural characteristics. According to the Cambridge Structural Database (CSD), more than 70,000 MOF structures have been designed and synthesised so far.[1] This figure is far higher than the number of reported zeolites which is limited to 64 natural structures and about 240 synthesised structures.[2] Due to the functional flexibility of their structures, MOFs possess exceptional properties such as apparent surface areas up to 10,000 $m^2 g^{-1}$ (compared to the maximum reported values of ~1,000 $m^2 g^{-1}$ for zeolites and 3,500 $m^2 g^{-1}$ for activated carbons[3,4]) and the highest porosity (3.60 $cm^3 g^{-1}$)[5] yet achieved for any porous material.[6] These unique properties lead to MOFs being trialled in various energy and environmentally relevant applications such as gas storage,[7–9] gas separation,[10–13] catalysis,[14,15] carbon dioxide capture[16–18] and as semiconductors.[19,20] Due to the enormous number of MOF entries dispersed throughout the literature and their potential for application, there is growing research effort focussed on the large-scale synthesis and practical application of novel MOF structures. Promising new approaches have focussed on novel synthetic methods such as supercritical fluid processing to reduce the environmental impacts associated with solvothermal synthesis[21–23] or mechanochemical synthesis to scale up the production of MOFs.[24] Other recent research strategies have involved tailoring porosity of conventional MOF-based structures to form hierarchically-structured porous materials,[23,25–27] to fabricate materials based on MOFs that can be tailored for use in numerous areas such as environmental remediation[28], catalysis,[29] energy,[30] health,[31] or in electronic devices and chemical sensors.[32]

Among these new synthetic approaches, the development of advanced hierarchical porous MOFs made of interconnected pores spanning different length scales ranging from micro-, to meso- and macroporosity (referring to pore diameters of <2 nm, 2-50 nm and >50 nm, respectively) has attracted considerable current interest. Introduction of larger meso- and macropores into traditionally microporous MOFs is desirable for facilitating diffusional processes, thereby increasing the rate of reaction for diffusion-limited or mass-transfer limited processes.[26,33] Numerous methods to synthesise these types of hierarchical porous MOFs have been developed, including template-assisted routes (e.g. dual surfactant templating methods[27,34] and $CO_2$-ionic liquid interfacial templating routes[35]), template-free routes (e.g. facile self-assembly processes[36] and use of $CO_2$-expanded solvents[23,37]), spontaneous formation (via linker labilization[38] and metal-ligand fragmentation[39–41]) and post-treatment (acid etching[42,43] and emulsion templating[44]).

While other methods require lengthy synthetic procedures to functionalise meso/macropores in hierarchical porous MOFs, acid etching stands out as the simplest method to create mesoporous defects and to engineer large voids via controlled acid etching. Recently, a process called size-selective acid diffusion was investigated and explained by Koo *et. al.*[45] By observing the etching process and the mesopore transformation of MIL-100(Fe), phosphoric acid with an appropriate size (d = 0.61 nm) was indicated to selectively diffuse into the 3D tetrahedral channel of MIL-100(Fe) through the hexagonal windows (d = 0.89 nm) rather than pentagonal windows (d = 0.49 nm) of the MOF. After gradual etching at 70 °C, the metal nodes and the benzene-1,3,5-tricarboxylic acid (BTC) linkers near this channel were disassembled, leading to the creation of large cages while maintaining the majority of the microporous cages, to preserve the crystallinity and external morphology. With the smaller acids, for example HCl (d = 0.34 nm), MIL-100(Fe) collapsed after etching because the small cages were also attacked, in contrast to the size-selective acid diffusion approach. A similar size-selective controlled etching process, termed "synergistic etching and surface functionalisation", was reported by Hu *et al.*[42] on a ZIF-8 MOF. In that research, weak bulky phenolic acids such as gallic acid and tannic acid were used to provide free protons that penetrating into the MOF crystals to form a hollow structure. Interestingly, these acids with large molecular size simultaneously blocked the exposed surface of ZIF-8, leading to the preservation of the parent crystalline framework of MOF in the outer shell. Typically, such use of acid etching to tune pore size in those processes is only applicable for highly water-stable MOFs. As the stability of MOFs is a recognised impediment to their industrial applicability and a large number of MOFs are not stable in water,[46,47] the generalisation of this inventive acid etching method remains a significant challenge.

Inspired by recent reports where phosphoric acid could be used to tailor porosity via selective acid etching, we report an innovative method to create hierarchical porous structure in water-unstable MOFs which has not been examined so far. Here, we used the archetypical HKUST-1 (Cu-BTC) MOF as a model. This study sought to investigate to what extent porosity may be tailorable by controlling the degree of acid etching in these systems.

## Results and discussion

Prior to etching, HKUST-1 was assessed for water stability at 40 °C to avoid unexpected degradation while controlling the degree of acid etching on HKUST-1. This was of importance for obtaining a defective hierarchical porous HKUST-1 while retaining the MOF microstructure. To investigate the degradation of HKUST-1 in water, 50 mg of normal HKUST-1 was immersed in 50 ml of distilled water for different lengths of time (from 1 to 72 h) at 40 °C without stirring. PXRD results of these samples show that HKUST-1 immersed in water for 1, 3 and 5 hours partially maintained its original crystallinity with the preservation of peaks at 7 and 12 degrees $2\theta$. However, after 24 h, this MOF started to noticeably change in structure with the absence of the peaks at 7 and 12 degrees $2\theta$ and an obvious split in the peak at 10 degrees $2\theta$ (Figure 1). Note that peaks at 14, 17 and 27 degrees $2\theta$ retained their positions in all samples, so the assumption of changes in unit cell parameter in the samples immersed in water for periods greater than 24 h can be disregarded. This result is consistent with what shown in the literature[48] where HKUST-1 was reported to be stable in relative humidity up to 90% at temperature up to 50 °C and for 6 hours. Disappearance of the characteristic diffraction peaks in PXRD results show that changing the concentration of the acid solution by using serial dilution in water (which was shown to be a very promising method for control of the degree of etching in MIL-100(Fe) from 2 to 12 hours in 40 mM of $H_3PO_4$[45]) would not be applicable to HKUST-1. In fact, HKUST-1 samples etched in a mixture of phosphoric acid and water at different concentration and time showed poor crystallinity and very low surface area (below 34 $m^2\ g^{-1}$), in comparison to normal HKUST-1 (more than 2,200 $m^2\ g^{-1}$) (Supplementary Figure S2, Table S2 and Table S3).

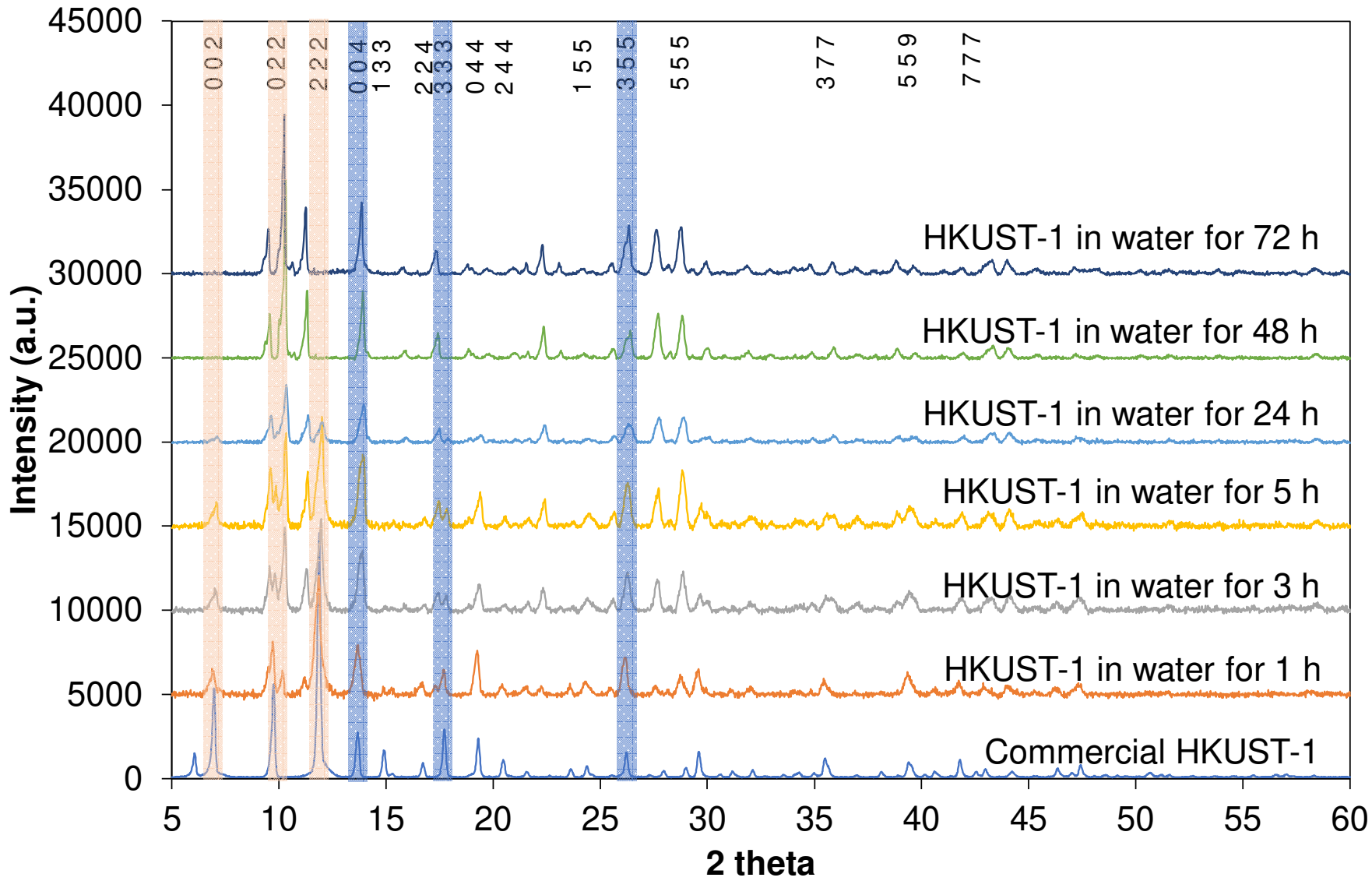

*Figure 1. PXRD results of HKUST-1 immersed in water from 1 to 72 hours. Key peaks that remained the same are highlighted in light blue. Key peaks that changed or disappeared are highlighted in light orange. PXRD spectra are offset in intensity, for clarity.*

It has been shown that HKUST-1 can be synthesised from a stable precursor solution formed by a mixture of trimesic acid, copper nitrate and DMSO[49]. Recently, we also reported a novel method to synthesise HKUST-1 using supercritical $CO_2$ to dramatically reduce the amount of MeOH needed to trigger the HKUST-1 nucleation reaction inside the precursor solution[23]. These results suggested that HKUST-1 would be stable in a mixture of DMSO and MeOH, indicating that this could be used to vary the concentration of the acid solution and control the degree of etching in HKUST-1. Indeed, the overall crystallinity of HKUST-1 remained unchanged with up to 10 days etching and up to an acidity of pH 2.6, which is confirmed by the series of PXRD results shown in Figure 2a and b. Note that the decrease in the intensity of the peak at 5.8 degrees $2\theta$ is because of different amounts of moisture adsorbed in the samples during XRD preparation. This was demonstrated by comparing XRD patterns of two simulated HKUST MOFs in hydrated and dehydrated form. It can be seen that the peak at 5.8 degrees $2\theta$ significantly decreased after removing all oxygen atoms of water molecules coordinated with copper in simulated dehydrated HKUST-1 structure (see Supplementary Figure S6).

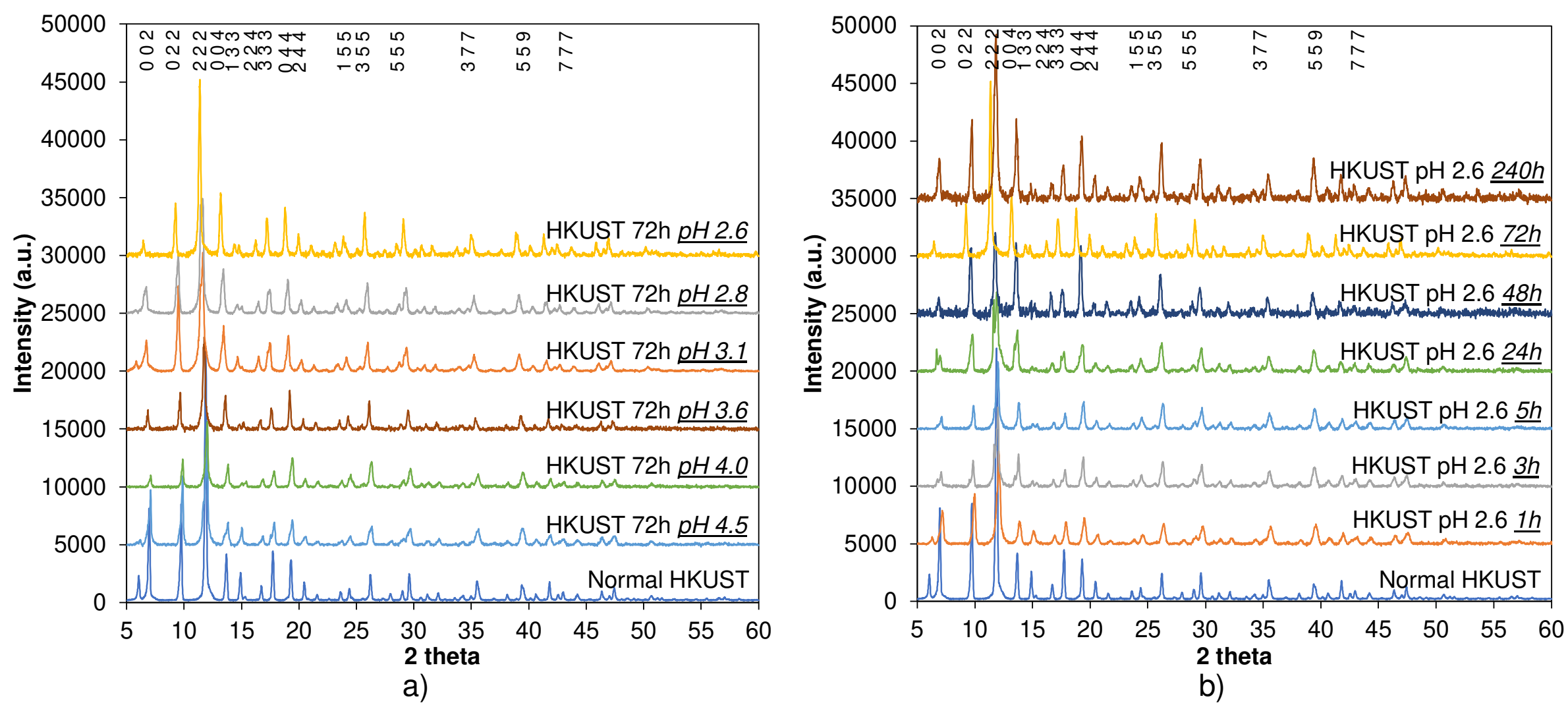

*Figure 2. PXRD results of HKUST-1 etching in phosphoric acid using DMSO and MeOH as dilute solvents at different times (a) and concentrations (b). PXRD spectra are offset in intensity, for clarity.*

To investigate the mass change after acid etching, the dry weights of the HKUST samples in centrifuge tubes were recorded before exposure to phosphoric acid and after washing with methanol 3 times to remove excess acid and drying the solid in an oven overnight at 60 °C. The acid etching of the HKUST-1 at pH 2.6 for 72 h was repeated 5 times at the same conditions to test reproducibility and to estimate experimental errors which could be caused by sample loss during the decanting of the solvents, drying or weighing. The deviation in the measured dry weights shows that the error bars in the measured weights for this experiment are about ±1.5 mg (Table S4 in SI), which was deemed reliable enough to initially investigate the etching process and compare between samples etched for different times and with different acid concentrations. Figure 3a shows that when etching at pH 2.6 for 72 h, the dried sample lost 24 wt%. The weight loss decreased when pH increased from 2.8 to 6.4, meaning that the lower the phosphoric acid concentration, the less weight loss (and presumably the less etching) was seen in the sample. When the pH was fixed at 2.6 and the etching time was varied from 1 h to 240 h, a reverse trend was observed. The weight loss of samples etched for short periods (1, 3 and 5 h) was between 13 and 18 wt%, and for longer periods of exposure to the acid (24, 48 and 72 h) the weight loss observed was between 22 and 24 wt%. The highest figure was seen in the sample etched for 240 h, reaching 42 wt% loss after etching in acid. From these results, it can be predicted that the degree of etching depends on both concentration and time.

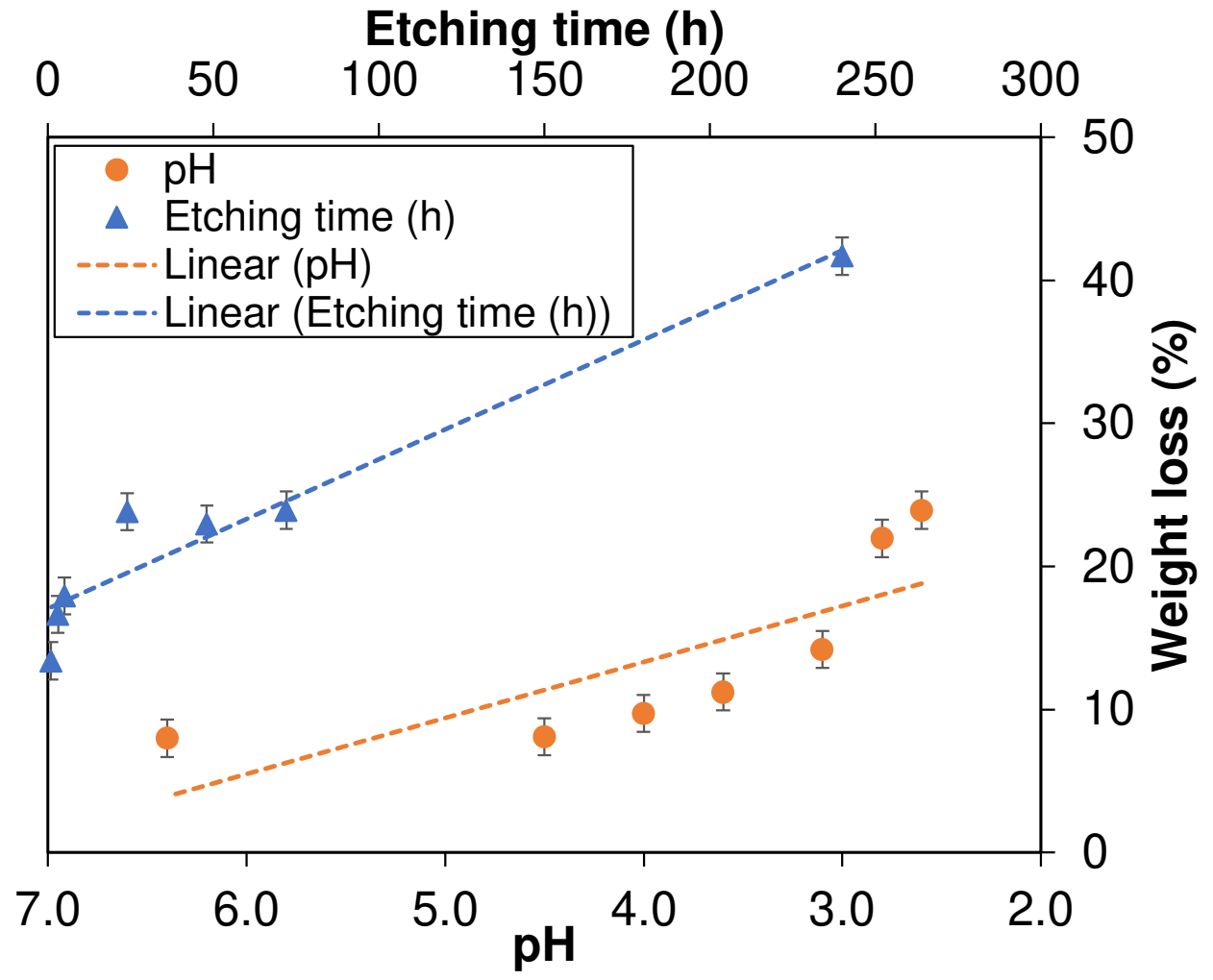

*Figure 3. Weight loss of HKUST-1 etching in phosphoric acid using DMSO and MeOH as dilute solvents at different concentrations and times. The dotted lines are linear fits, intended as guides to the eye.*

Samples were examined by SEM to investigate the effects of etching on crystal morphology and surface structure. The untreated HKUST crystals were faceted and typically up to 40 microns in diameter (Figure 4a). These features were retained on exposure to the pH 6.4 DMSO/MeOH stock solution (Figure 4b). There was no discernible etching by the SS of the exposed crystal faces (see insets Figures 4a and 4b). In contrast, for the samples etched in phosphoric acid for a constant etching time of 72 h using DMSO and MeOH as dilute solvents at different acidities ranging between 2.6 and 4.5 (Figure 4c-h), it can be seen that at less acidic conditions (pH at 3.6 and above) there was no appearance of macropores in these samples even though some features witnessed on the surface of HKUST 72h pH 3.6 show some minor etching effects which can be evidenced in the SEM images in Figure 4 c and e. At higher concentrations of phosphoric acid in DMSO and MeOH, interestingly, a number of large hexagonal pores with an approximate diameter size of 0.5 μm can be observed. The number of these newly-introduced holes seemed to increase while increasing the acid concentration. Surprisingly, the regular, well-ordered hexagonal shape of these holes in HKUST-1 has not been reported even though this MOF is a favoured candidate for study of porosity tailoring and surface functionalisation.

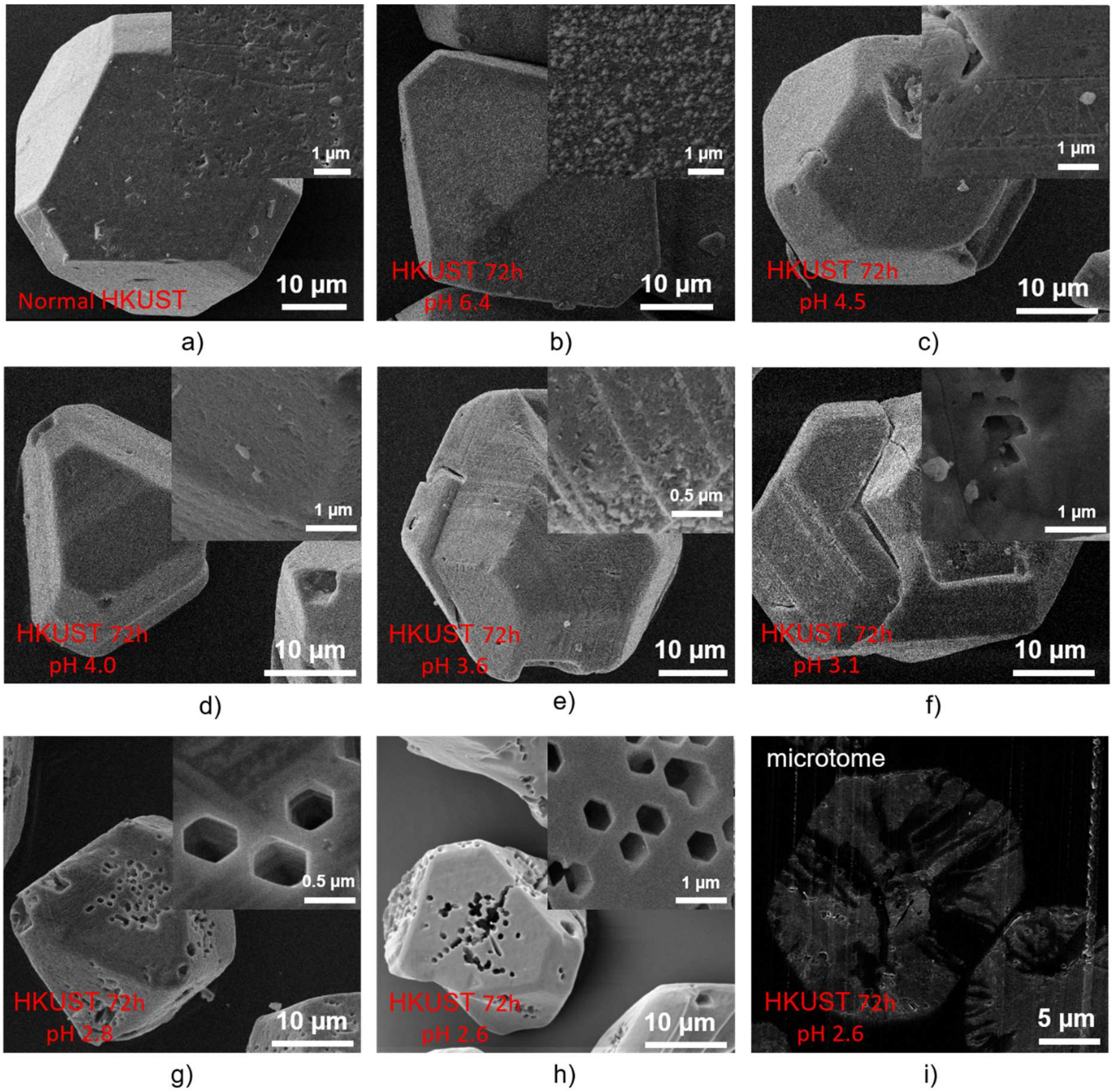

*Figure 4. SEM images of HKUST-1 etching in phosphoric acid using DMSO and MeOH as dilute solvents at different concentrations.*

To examine the effect of etching time on creation of these macropores in HKUST-1, the acidity was maintained at 2.6 (1 ml $H_3PO_4$ in 10 ml DMSO and 10 ml MeOH) and the etching time was varied from 1 hour to 240 hours (Figure 5a-f). Small holes with some triangle-shaped geometrical edge features appeared on the surface of the sample etching for 1 hour (Figure 5a), which may suggest the direction of creating hexagonal pores observed previously in the sample etching in 72 hours (Figure 4h). It is also notable that the large hexagonal pores did not appear when using a shorter etching time (up to 5 hours; Figure 5a-c) but appeared when increasing etching time to 24 hours and above (Figure 5d-f). These results are consistent with the non-linear time-dependent porosity generation with supercritical $CO_2$ exposure time we reported previously.[23] The number of these hexagonal holes in the sample etched over 240 h (Figure 5f) is obviously higher than those in the samples etched over 24 h (Figure 5d), 48 h (Figure 5e) and 72 h (Figure 4h). It can also be seen that the crystal morphology of the original HKUST-1 samples was still preserved (compared with Figure 4a) with an octahedral shape and a crystal size of 15-20 μm. The sample etching only in DMSO and MeOH (pH = 6.4) in 72 h do not result in any noticeable changes on the surface (Figure 4b), indicating that phosphoric acid played a critical role in creating these hexagonal holes. To show the inner connectivity of these holes, the sample that was etched at acidity of pH 2.6 over 72 h was carefully sectioned using a microtome and observed again under the SEM. Interestingly, the etching in phosphoric acid using DMSO and MeOH as dilute solvents not only generated hexagonal macropores while preserving the HKUST-1 microstructure and external crystal morphology but also prompted formation of a connected interior pore network between these holes (Figure 4i), which might be very useful to improve molecular diffusion in catalytic processes.

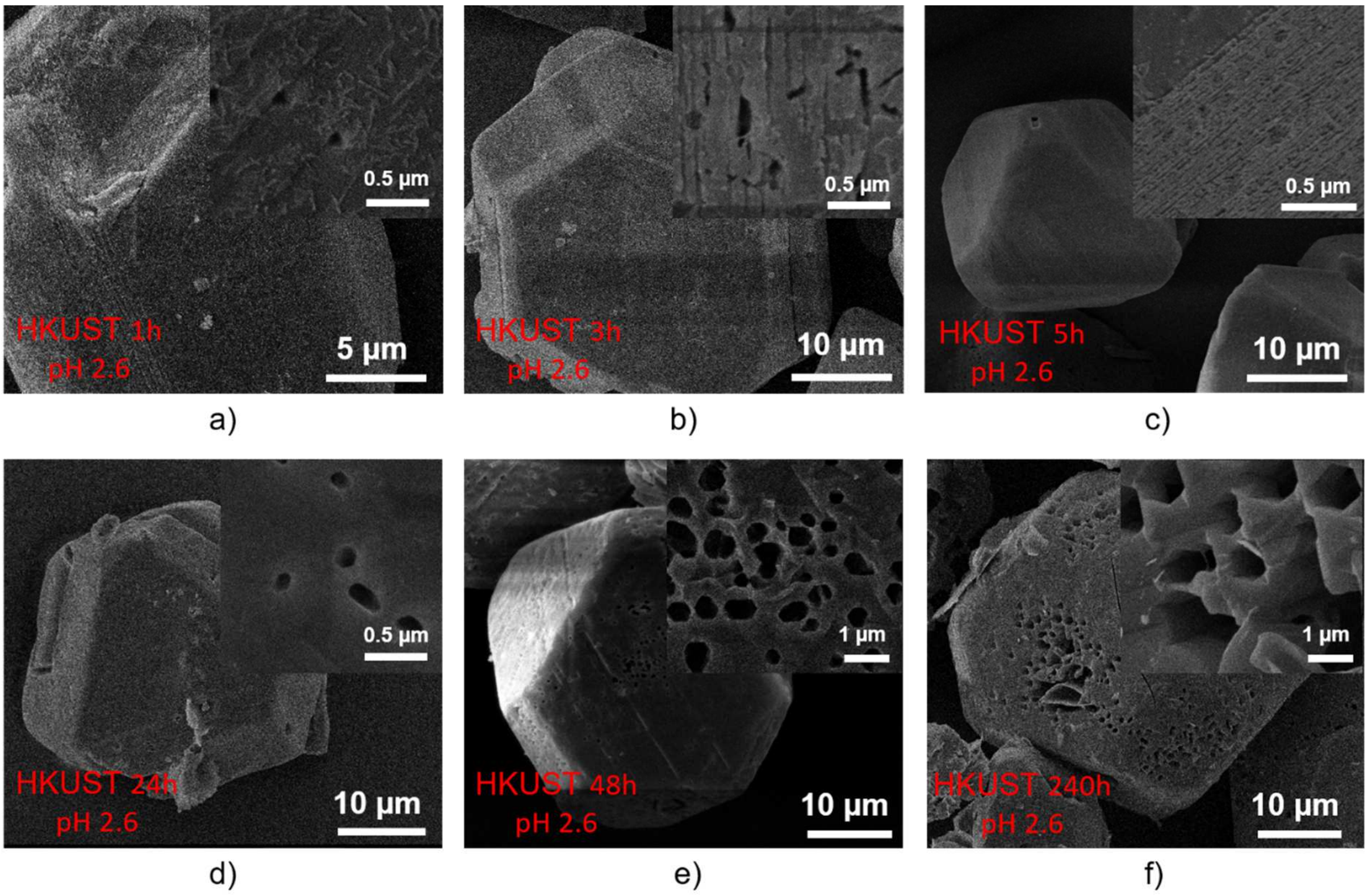

*Figure 5. SEM images of HKUST-1 etching in phosphoric acid using DMSO and MeOH as dilute solvents at different times.*

The surface areas and total pore volumes of each of the samples were examined by using $N_2$ sorption at 77 K. In the isotherm data shown in Figure 6, all etched samples possess the typical microporous isotherms with two steps clearly observed in the low relative pressure region ($P/P_o$ below than $10^{-3}$). This adsorption isotherm is to be expected in normal HKUST-1 which has larger primary pores (d = 0.83 nm) and smaller secondary pores (d = 0.33 nm). This result confirms that the micropore structure within the samples was maintained after etching, in contrast to the distinct loss of microporosity when etching HKUST-1 in phosphoric acid using water as the key solvent (see Supplementary Table S2 and S3). When comparing the samples etched in different concentrations as well as for different times, a small but steady decrease in both surface areas and total pore volume was observed for higher pH and with longer times in the phosphoric acid stock solution (Figure 7). This is consistent with the weight loss data discussed previously, demonstrating that the more weight loss the lower the surface area and total pore volume in the etched samples. The lowest surface area and total pore volume were seen in the HKUST-1 etching in pH 2.6 for 240 h (1,117 $cm^2\ g^{-1}$ and 0.52 $cm^3\ g^{-1}$) which are significantly lower values than normal HKUST-1 (2,208 $cm^2\ g^{-1}$ and 0.90 $cm^3\ g^{-1}$). This is due to a greater proportion of macropores (about 0.5 μm) in those samples, which cannot be analysed by nitrogen adsorption/desorption isotherms. Similar trends were also shown in HKUST-1 synthesised in supercritical $CO_2$[23] and UiO-66 in acid etching[50].

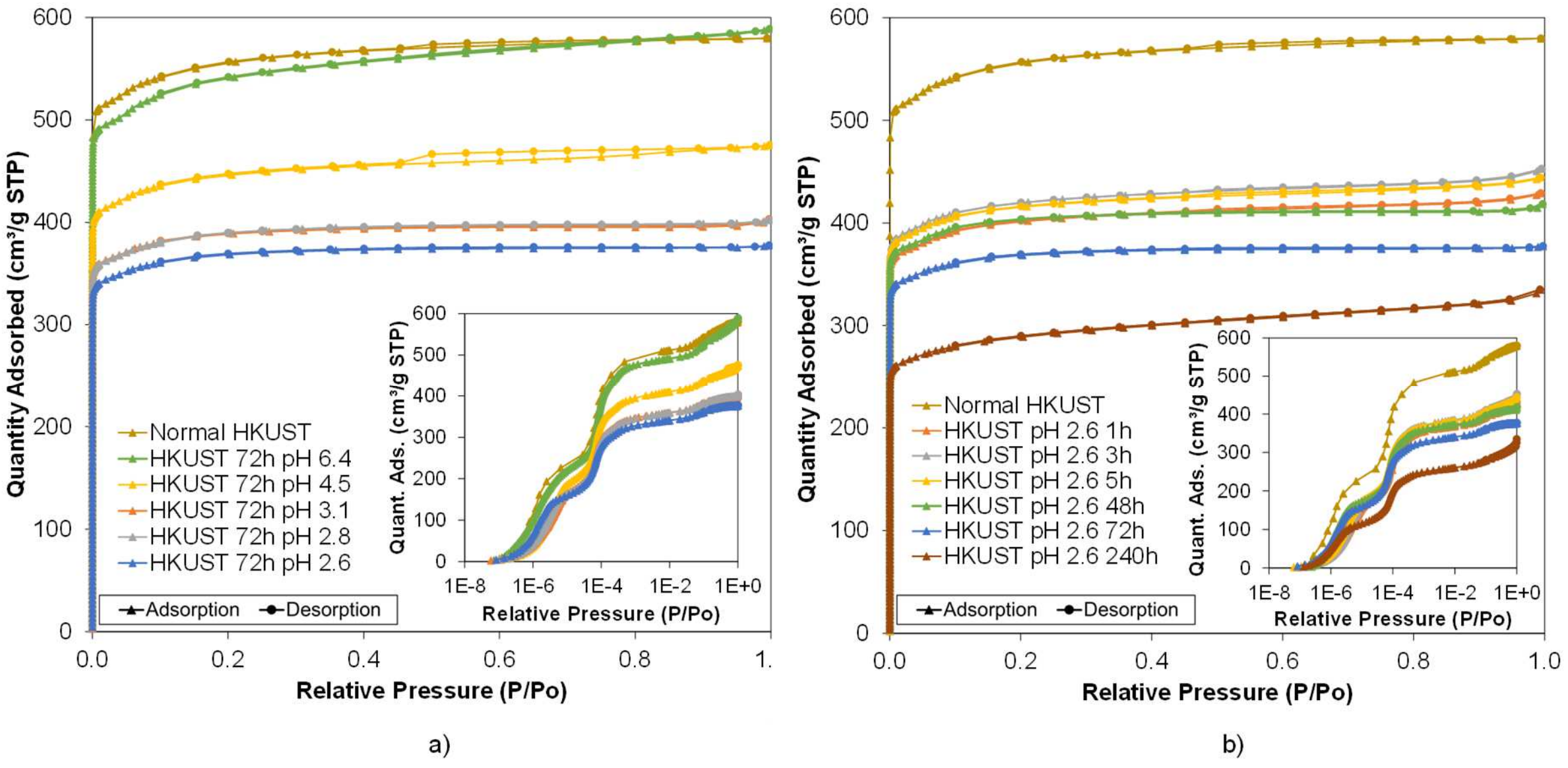

*Figure 6. Nitrogen isotherms of HKUST-1 etching in phosphoric acid using DMSO and MeOH as dilute solvents at different concentrations (a) and times (b). The full isotherms in logarithmic scale in the blown-up sections show 2-stepped adsorption in the samples, which correspond to two different micropores preserved after etching.*

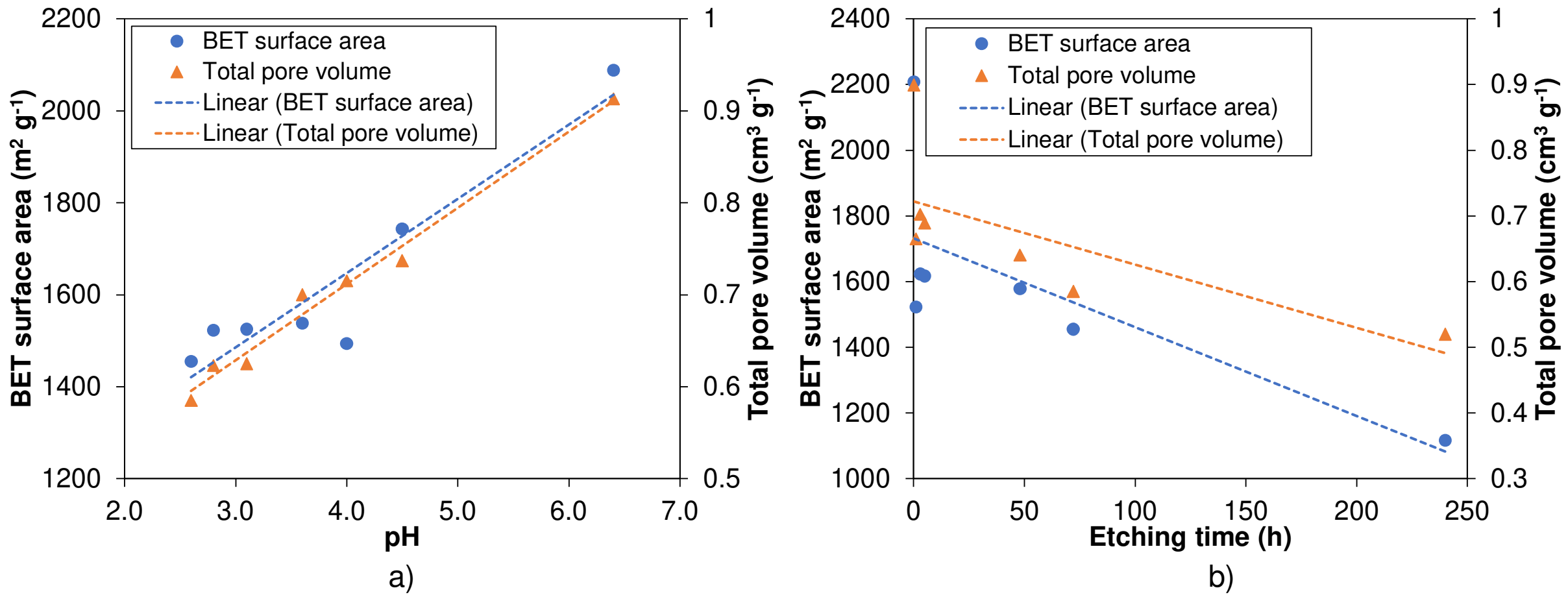

*Figure 7. Gas sorption results of HKUST-1 etching in phosphoric acid using DMSO and MeOH as dilute solvents at different concentrations (a) and times (b). The dotted lines are linear fits, intended as guides to the eye.*

The disassembly of 1 copper cluster and 4 linkers (paddlewheels) was considered the main reason of defect formation in HKUST-1.[40,41] This defect formation was confirmed by TGA, which indicated by proportionally lower weight loss over the region between 250 °C and 300 °C for the etched acid- samples (see Figure S4 in Supplementary Information ). $^{31}$P NMR additionally confirmed the defect formation did not result in coordination of $PO_4^{3-}$ in the MOF structure (Figure S5 in Supplementary Information). This cluster disassembly mechanism can be used to explain the formation of large hexagonal cages, which would be, in turn, more susceptible to selective acid etching in HKUST-1 using $H_3PO_4$ (Figure 8b). Removal of some secondary building units can also be used to explain the lower BET surface areas after HKUST-1 acid etching (Figure 7). The longer the HKUST-1 was exposed to etching in $H_3PO_4$, a greater number of clusters and linkers in the tetrahedral channel were disassembled, resulting in a greater number of large hexagonal voids being observed. Remarkably, the overall crystallinity was still preserved due to the maintained structural integrity formed by the majority of secondary building units in the small cages which were not easily accessible by the acid and therefore not affected by the selective acid etching.

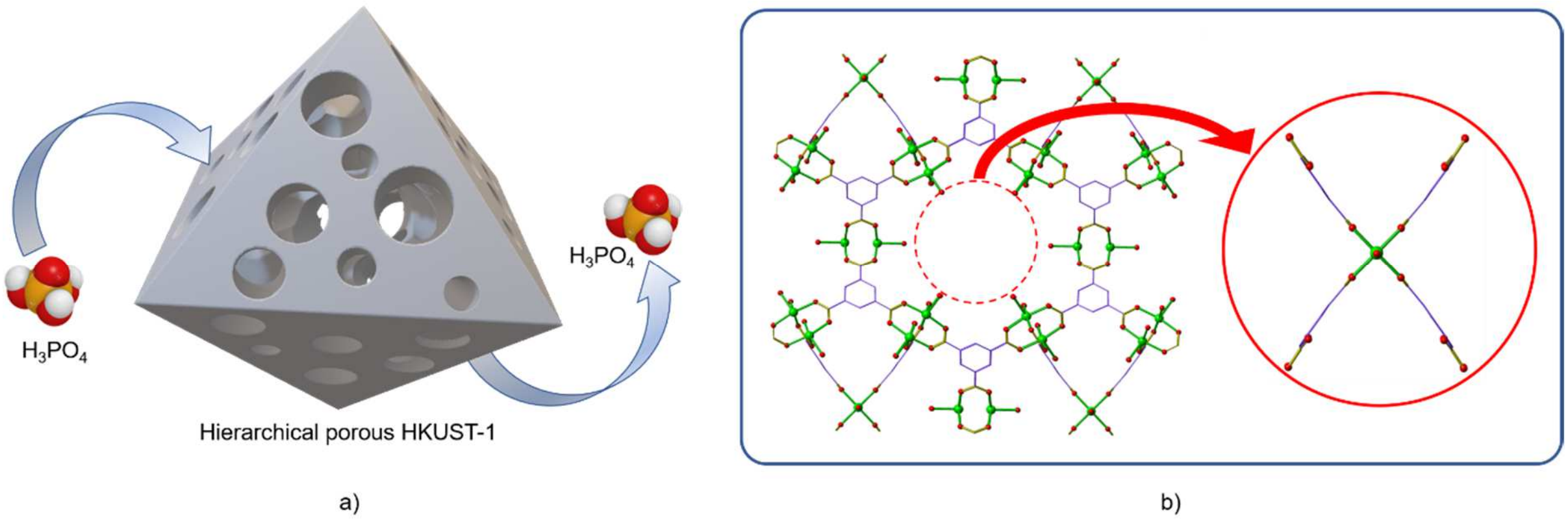

*Figure 8. Etching process for HKUST-1 using phosphoric acid. a) Phosphoric acid diffuses into HKUST-1 to form a hierarchical porous structure. b) The disassembly of a cluster and 4 linkers.*

## Conclusion

In this study, an innovative approach to acid etching was systematically investigated on a MOF (HKUST-1) that is known not to be stable in water. It is demonstrated that the extent of etching on this MOF depends on both pH and etching time. Geometrical pores were observed on HKUST-1 MOF immersed in phosphoric acid, DMSO and methanol at pH less than 2.8 for more than 24 h. It is predicted that by using phosphoric acid with a size-selective diameter (d = 0.61 nm), the diffusion process occurs preferably in larger pores (d = 0.83 nm) rather than in smaller pores (d = 0.33 nm) of HKUST-1, resulting in the interconnected macropores appearing over longer etching times and higher acid concentrations while the overall crystallinity, morphology and microporosity of this MOF was still preserved. The tailorable hierarchical pore structure obtained by this method would be promising for the use of MOF catalysts, to speed up reactions of bulky molecules where the rate of reaction is limited by low diffusion process or high pressure drop.

## Materials and Methods

All chemicals used in this study were purchased from commercial suppliers without further purification (see Supplementary Table S1). A commercial HKUST-1 (also known as Basolite C 300, copper benzene-1,3,5-tricarboxylate or Cu-BTC and referred here as normal HKUST) was sourced from Sigma-Aldrich and was subjected to acid etching, as follows. To test the effect of pH on the etching process, 50 ml DMSO and 50 ml methanol were mixed and magnetically stirred at room temperature for 10 minutes to obtain a stock solution (SS) with

pH = 6.4. Subsequently, different amounts of 90% phosphoric acid (0.2 ml, 0.6 ml and 1 ml) were added to 20 ml of the stock solution, with continuous stirring at room temperature for another 10 minutes. The acidities of these solutions were recorded by a calibrated Jenway 3510 pH meter with an accuracy of ±0.003. The stock solution containing 1 ml phosphoric acid is subsequently referred to as SSH. To increase the pH, 20 ml of SS was acidified by addition of 0.2 ml, 0.6 ml and 1 ml of SSH, with vigorous stirring before measuring pH (see Table 1). In the etching process, 20 ml of the prepared solution with different pH from 2.6 to 6.4 was added to 50 ml centrifuge tubes containing 200 mg of normal HKUST, and was agitated strongly before being placed in an oven at 40 °C for 72 h. To evaluate the effect of etching time, the same amount of normal HKUST was soaked in 20 ml SSH at 40 °C for 1 h, 3 h, 5 h, 24 h, 48 h, 72 h and 240 h. The details of the HKUST-1 acid etching process are summarised in Supplementary Figure S1 After etching, the solid was isolated by centrifugation (10,000 rpm for 10 minutes), washed with pure methanol, re-centrifuged and re-washed three times and dried at 60 °C overnight. The dried samples in the centrifuge tubes were weighed before and after etching to record the weight loss with respect to the starting dry weight.

*Table 1. Preparation of stock solutions (a mixture of phosphoric acid, DMSO and methanol) with different pH for testing HKUST in acid etching.*

| | **Stock solution (SS)** | **SS + 1 ml $H_3PO_4$ (SSH)** | **SS + 0.6 ml $H_3PO_4$** | **SS + 0.2 ml $H_3PO_4$** | **SS + 1 ml SSH** | **SS + 0.6 ml SSH** | **SS + 0.2 ml SSH** |
|---|---|---|---|---|---|---|---|
| **pH** | 6.4 | 2.6 | 2.8 | 3.1 | 3.6 | 4.0 | 4.5 |

The etched HKUST MOFs were characterised by powder X-Ray diffraction (PXRD) on a BRUKER AXS D8-Advance instrument (Cu Kα radiation, λ = 1.5418 Å, in flat plate geometry) to check the crystallinity and phase, thermogravimetric analysis (TGA) at 5 °C $min^{-1}$ ramp rate in air to determine changes in the metal-linker ratios with etching, nitrogen sorption analysis at 77 K using a Micrometrics 3 Flex (degassing at 120 °C under dynamic high vacuum over 6 hours using at least 100 mg of sample) to measure BET surface area and pore size distribution and scanning electron microscopy (SEM) using a JSM-IT300 instrument with a secondary electron detector (coating the samples with 10 nm silver) to investigate crystal morphology and macroporosity. In addition, a representative HKUST single crystal etched in pH 2.6 for 72 h was embedded in an epoxy resin, cured at 60 °C for 12 h, and sectioned using an ultra-microtome to image its internal structure. The microtomed block was coated with a thin layer of silver (10 nm) and imaged with SEM at 5 kV using a back-scattered electron detector to reveal the pore connectivity. HKUST samples were dissolved in 0.1 ml deuterium chloride (DCl, 99 atom% D) and 3 ml anhydrous dimethyl sulfoxide-$d_6$ (DMSO-$d_6$, 99 atom% D) before being tested by $^{31}P$ nuclear magnetic resonance (NMR). Further details of these experiments are provided in Supplementary Information.

***References***

1. Moghadam, P. Z. *et al.* Development of a Cambridge Structural Database Subset: A Collection of Metal–Organic Frameworks for Past, Present, and Future. *Chem. Mater.* **29,** 2618–2625 (2017).
2. Ch. Baerlocher and L.B. McCusker. Database of Zeolite Structures. Available at: http://www.iza-structure.org/databases/.
3. Farha, O. K. *et al.* Metal–Organic Framework Materials with Ultrahigh Surface Areas: Is the Sky the Limit? *J. Am. Chem. Soc.* **134,** 15016–15021 (2012).
4. Gómez-Gualdrón, D. A., Moghadam, P. Z., Hupp, J. T., Farha, O. K. & Snurr, R. Q. Application of Consistency Criteria to Calculate BET Areas of Micro- and Mesoporous Metal-Organic Frameworks. *J. Am. Chem. Soc.* **138,** 215–224 (2016).
5. Farha, O. K. *et al.* De novo synthesis of a metal–organic framework material featuring ultrahigh surface area and gas storage capacities. *Nat. Chem.* **2,** 944–948 (2010).
6. Furukawa, H. *et al.* Ultrahigh porosity in metal-organic frameworks. *Science (80-. ).* **329,** 424–428 (2010).
7. Gó Mez-Gualdró, D. A. *et al.* Evaluating topologically diverse metal-organic frameworks for cryo-adsorbed hydrogen storage †. *Energy Environ. Sci* **9,** 49 (2016).
8. Mason, J. A. *et al.* Methane storage in flexible metal–organic frameworks with intrinsic thermal management. *Nature* **527,** 357–361 (2015).
9. Suh, M. P., Park, H. J., Prasad, T. K. & Lim, D.-W. Hydrogen Storage in Metal–Organic Frameworks. *Chem. Rev.* **112,** 782–835 (2012).
10. Holcroft, J. M. *et al.* Carbohydrate-Mediated Purification of Petrochemicals. *J. Am. Chem. Soc.* **137,** 5706–5719 (2015).

11. Li, J.-R., Kuppler, R. J. & Zhou, H.-C. Selective gas adsorption and separation in metal–organic frameworks. *Chem. Soc. Rev.* **38,** 1477 (2009).

12. DeCoste, J. B. & Peterson, G. W. Metal–Organic Frameworks for Air Purification of Toxic Chemicals. *Chem. Rev.* **114,** 5695–5727 (2014).

13. Li, J.-R., Sculley, J. & Zhou, H.-C. Metal–Organic Frameworks for Separations. *Chem. Rev.* **112,** 869–932 (2012).

14. Liu, J. *et al.* Applications of metal–organic frameworks in heterogeneous supramolecular catalysis. *Chem. Soc. Rev* **43,** (2014).

15. Corma, A., García, H. & Llabrés i Xamena, F. X. Engineering metal organic frameworks for heterogeneus catalysis. *Chem. Rev.* **110,** 4606–4655 (2010).

16. Sumida, K. *et al.* Carbon Dioxide Capture in Metal–Organic Frameworks. *Chem. Rev.* **112,** 724–781 (2012).

17. McDonald, T. M. *et al.* Cooperative insertion of $CO_2$ in diamine-appended metal-organic frameworks. *Nature* **519,** 303–308 (2015).

18. Yazaydın, A. O. *et al.* Screening of metal– organic frameworks for carbon dioxide capture from flue gas using a combined experimental and modeling approach. *J. Am. Chem. Soc.* **131,** 18198–18199 (2009).

19. Férey, G. Hybrid porous solids: Past, present, future. *Chem. Soc. Rev.* **37,** 191–214 (2008).

20. Sheberla, D. *et al.* High Electrical Conductivity in $Ni_3$(2,3,6,7,10,11-hexaiminotriphenylene)$_2$, a Semiconducting Metal–Organic Graphene Analogue. *J. Am. Chem. Soc.* **136,** 8859–8862 (2014).

21. Matsuyama, K. Supercritical fluid processing for metal–organic frameworks, porous coordination polymers, and covalent organic frameworks. *J. Supercrit. Fluids* **134,** 197–203 (2018).

22. López-Periago, A. M. & Domingo, C. Features of supercritical $CO_2$ in the delicate world of the nanopores. *J. Supercrit. Fluids* **134,** 204–213 (2018).

23. Doan, H. V. *et al.* Controlled Formation of Hierarchical Metal–Organic Frameworks Using $CO_2$-Expanded Solvent Systems. *ACS Sustain. Chem. Eng.* **5,** 7887–7893 (2017).

24. Rubio-Martinez, M. *et al.* New synthetic routes towards MOF production at scale. *Chem. Soc. Rev.* **46,** 3453–3480 (2017).

25. Yang, X.-Y. *et al.* Hierarchically porous materials: synthesis strategies and structure design. *Chem. Soc. Rev.* **46,** 481–558 (2017).

26. Duan, C. *et al.* Facile synthesis of hierarchical porous metal-organic frameworks with enhanced catalytic activity. *Chem. Eng. J.* **334,** 1477–1483 (2018).

27. Duan, C. *et al.* Template synthesis of hierarchical porous metal–organic frameworks with tunable porosity. *RSC Adv.* **7,** 52245–52251 (2017).

28. Bobbitt, N. S. *et al.* Metal–organic frameworks for the removal of toxic industrial chemicals and chemical warfare agents. *Chem. Soc. Rev.* **46,** 3357–3385 (2017).

29. Rogge, S. M. J. *et al.* Metal–organic and covalent organic frameworks as single-site catalysts. *Chem. Soc. Rev.* **46,** 3134–3184 (2017).

30. Adil, K. *et al.* Gas/vapour separation using ultra-microporous metal–organic frameworks: insights into the structure/separation relationship. *Chem. Soc. Rev.* **46,** 3402–3430 (2017).

31. Lian, X. *et al.* Enzyme–MOF (metal–organic framework) composites. *Chem. Soc. Rev.* **46,** 3386–3401 (2017).

32. Lustig, W. P. *et al.* Metal–organic frameworks: functional luminescent and photonic materials for sensing applications. *Chem. Soc. Rev.* **46,** 3242–3285 (2017).

33. Parlett, C. M. A., Wilson, K. & Lee, A. F. Hierarchical porous materials: catalytic applications. *Chem. Soc. Rev.* **42,** 3876–3893 (2013).

34. Duan, C. *et al.* Synthesis of Hierarchically Structured Metal−Organic Frameworks by a Dual-Functional Surfactant. *ChemistrySelect* **3,** 5313–5320 (2018).

35. Peng, L. *et al.* Hollow metal–organic framework polyhedra synthesized by a $CO_2$–ionic liquid interfacial templating route. *J. Colloid Interface Sci.* **416,** 198–204 (2014).

36. Yang, X., Wu, S., Wang, P. & Yang, L. Hierarchical 3D ordered meso-/macroporous metal-organic framework produced through a facile template-free self-assembly. *J. Solid State Chem.* **258,** 220–224 (2018).

37. Peng, L. *et al.* Highly mesoporous metal–organic framework assembled in a switchable solvent. *Nat. Commun.* **5,** 933–969 (2014).

38. Yuan, S. *et al.* Construction of hierarchically porous metal–organic frameworks through linker labilization. *Nat. Commun.* **8,** 15356 (2017).

39. Park, J., Wang, Z. U., Sun, L.-B., Chen, Y.-P. & Zhou, H.-C. Introduction of Functionalized Mesopores to Metal–Organic Frameworks via Metal–Ligand–Fragment Coassembly. *J. Am. Chem. Soc.* **134,** 20110–20116 (2012).

40. Kim, S.-Y., Kim, A.-R., Yoon, J. W., Kim, H.-J. & Bae, Y.-S. Creation of mesoporous defects in a microporous metal-organic framework by an acetic acid-fragmented linker co-assembly and its remarkable effects on methane uptake. *Chem. Eng. J.* **335,** 94–100 (2018).

41. Zhang, W. *et al.* Impact of Synthesis Parameters on the Formation of Defects in HKUST-1. *Eur. J. Inorg. Chem.* **2017,** 925–931 (2017).

42. Hu, M. *et al.* Void Engineering in Metal-Organic Frameworks via Synergistic Etching and Surface Functionalization. *Adv. Funct. Mater.* **26,** 5827–5834 (2016).

43. El-Hankari, S., Huo, J., Ahmed, A., Zhang, H. & Bradshaw, D. Surface etching of HKUST-1 promoted via supramolecular interactions for chromatography. *J. Mater. Chem. A* **2,** 13479–13485 (2014).

44. Zhang, B. *et al.* High-internal-phase emulsions stabilized by metal-organic frameworks and derivation of ultralight metal-organic aerogels. *Sci. Rep.* **6,** 21401 (2016).

45. Koo, J. *et al.* Hollowing out MOFs: Hierarchical micro- and mesoporous MOFs with tailorable porosity via selective acid etching. *Chem. Sci.* **8,** 6799–6803 (2017).

46. Howarth, A. J. *et al.* Chemical, thermal and mechanical stabilities of metal–organic frameworks. *Nat. Rev. Mater.* **1,** 15018 (2016).

47. Burtch, N. C., Jasuja, H. & Walton, K. S. Water Stability and Adsorption in Metal–Organic Frameworks. *Chem. Rev.* **114,** 10575–10612 (2014).

48. Al-Janabi, N. *et al.* Mapping the Cu-BTC metal-organic framework (HKUST-1) stability envelope in the presence of water vapour for $CO_2$ adsorption from flue gases. *Chem. Eng. J.* **281,** 669–677 (2015).

49. Ameloot, R. *et al.* Direct patterning of oriented metal-organic framework crystals via control over crystallization kinetics in clear precursor solutions. *Adv. Mater.* **22,** 2685–8 (2010).

50. Dissegna, S. *et al.* Tuning the Mechanical Response of Metal–Organic Frameworks by Defect Engineering. *J. Am. Chem. Soc.* **140,** 11581–11584 (2018).

## Acknowledgements

HVD thanks the Vietnamese Government (911 scholarship) and the University of Bristol (Queen's School studentship) for the funding to support this research. VPT acknowledges support from the UK Engineering and Physical Sciences Research Council (EP/R01650X/1). SEM studies were carried out in the Chemical Imaging Facility, University of Bristol with equipment funded by EPSRC under Grant "Atoms to Applications" (EP/K035746/1). We thank Dr Rémi Castaing (Material and Chemical Characterisation Facility, University of Bath) for help with the TGA experiments, Dr Ulrich Hintermair (University of Bath) for useful discussions on the etching process, and Dr Harina Amer Hamzah (University of Bristol) for help with the NMR analysis.

## Author contributions

HVD carried out the acid etching experiments, PXRD, TGA, SEM, gas sorption measurements, data analysis and prepared the manuscript. VPT provided advice and consultation on the experiments, data analysis and manuscript preparation. AS provided advice on PXRD analysis of HKUST-1 after etching. JCE and SD performed the SEM of microtomed HKUST-1. All authors discussed the results and implications and commented on the manuscript.

## Author information

### Affiliations

*Department of Mechanical Engineering, University of Bristol, Bristol BS8 1TH, UK*

Huan V. Doan and Valeska P. Ting

*Department of Oil Refining and Petrochemistry, Faculty of Oil and Gas, Hanoi University of Mining and Geology, Duc Thang, Bac Tu Liem, Hanoi, Vietnam*

Huan V. Doan

*Department of Chemistry, University of Bath, Claverton Down, Bath, BA2 7AY, UK.*

Asel Sartbaeva

*Chemical Imaging Facility, School of Chemistry, University of Bristol, Bristol BS8 1TS, UK*

Jean-Charles Eloi

*School of Chemistry, University of Bristol, Bristol BS8 1TS, UK*

Sean Davis

**Competing interests:** The authors declare no competing interest.